\documentclass[10pt, conference, compsocconf]{IEEEtran}

\usepackage{booktabs}
\usepackage{listings}
\usepackage{array}
\usepackage{cite}
\usepackage[normalem]{ulem}
\usepackage{xcolor}
\usepackage[hidelinks]{hyperref}
\definecolor{urls}{rgb}{0.1,0.4,0.6}
\newcommand{\link}[1]{\leavevmode\color{urls}\href{https://#1}{#1}}

\newcommand{\ii}[1]{\emph{1.#1}}
\newcommand{\ip}[1]{\emph{2.#1}}

\newcommand{\category}[1]{
  \subsubsection{#1}}

\newcounter{OBS}

\newcommand{\obs}[2]{
  \refstepcounter{OBS}
  \emph{Observation \arabic{OBS}:} \emph{#2}\label{#1}}
  
\newcommand{\impl}[1]{
  \subsection{#1}}

\title{Failures and Fixes: A Study of Software System Incident Response}

\author{
\IEEEauthorblockN{
  Jonathan Sillito, Esdras Kutomi
}
\IEEEauthorblockA{
  Department of Computer Science, Brigham Young University, Provo, USA \\
  sillito@byu.edu, esdras.kutomi@gmail.com}
}

\begin{document}
\maketitle

\begin{abstract}
This paper presents the results of a research study related to software system failures, with the goal of understanding how we might better evolve, maintain and support software systems in production. We have qualitatively analyzed thirty incidents: fifteen collected through in depth interviews with engineers, and fifteen sampled from publicly published incident reports (generally produced as part of postmortem reviews). Our analysis focused on understanding and categorizing how failures occurred, and how they were detected, investigated and mitigated. We also captured analytic insights related to the current state of the practice and associated challenges in the form of \ref{obs:careful} key observations. For example, we observed that failures can cascade through a system leading to major outages; and that often engineers do not understand the scaling limits of systems they are supporting until those limits are exceeded. We argue that the challenges we have identified can lead to improvements to how systems are engineered and supported.
\end{abstract}


\section{Introduction}

When a software system experiences an outage or is degraded in functionality or performance,  engineers (going by various titles in different organizations~\cite{google:sre:2016}) are notified to investigate and mitigate the problem. In this paper, we refer to such an event as an \emph{incident} and we refer to the work of engineers in this context as \emph{incident response}. After an incident is mitigated, organizations may conduct a postmortem analysis of the incident and produce an \emph{incident report}~\cite{collier:1996}. Incidents and incident response activities have not been widely studied in academia, but they represent important software engineering work that is particularly relevant in situations where systems already in production are maintained and evolved, and an outage or degradation must be addressed quickly due to potential financial or other costs.

Suppose hypothetically that the availability of a system (i.e., the proportion of requests to the system that succeed) drops below some designated level and the monitoring system notifies an on-call engineer. The engineer then uses a range of tools to investigate the cause of the drop. In this example scenario, suppose she finds that the drop in availability is correlated with the deployment of a new version of some component of the system, and so she initiates a rollback of that deployment as a mitigation strategy. The engineer then continues to monitor the system using the same tools she used to investigate the problem. Once the problem has been mitigated, the incident response (as we are defining it) is complete, though additional work may be needed to identify and fix the defect so that the release can be redeployed.

Work in multiple software engineering research areas can be seen as (at least partially) aiming to reduce the likelihood of incidents occurring as software that is already in production is evolved (see for example work on defect detection and prediction~\cite{huang:icsme:2017}). Other work aims to reduce the time it takes to mitigate incidents when they do occur, often focusing on improving the way we design the infrastructure tools used to monitor software systems and investigate system state and behavior (see for example work on debugging and execution traces~\cite{shimari:icsme:2019}). Both the goal of preventing incidents and the goal of reducing time to mitigation will benefit from a deeper understanding of both incidents and incident response processes, as they are experienced by engineers today. To this end, this paper reports on a qualitative analysis of thirty incidents: fifteen collected from in depth interviews with software engineers who have incident response experience, and fifteen sampled from publicly published incident reports. Our analytic results provide insights into how incidents occur; and are detected, investigated and mitigated; all of which has implications for supporting software systems.



\section{Related Work}

What follows immediately below is a discussion of related work to help further situate the present work. Some related work is discussed later in the findings and implications sections (sections~\ref{results} and~\ref{implications}), allowing us to present it along with the results of our analysis to which it is related.

Our work is in the spirit of previous work that has used qualitative methods, such as interviews, observation and document analysis, to develop an understanding of particular software engineering tasks. For example, previous work has investigated programming information needs~\cite{ko:icse:2007}, source code investigation~\cite{sillito:tse:2008}, debugging~\cite{layman:esem:2013}, software release tasks~\cite{phillips:cscw:2012} and cross-disciplinary collaborative work~\cite{li:chase:2017}. While we are using similar data collection and analysis techniques, the work reported in this paper is aimed at improving our understanding of incident response activities, which have not yet been studied in this way. As a result, our work focuses on the state of the \emph{practice}, more than the state of the \emph{research}.

Incident response work is done by people in different roles and with different titles, depending on the organization: software engineers, system administrators, site reliability engineers, and likely others as well. Beyer, et al. describe the work of site reliability engineers, and include discussions of some of the topics we have studied (most notably system monitoring and responding to incidents) from the perspective of one organization~\cite{google:sre:2016}. Field studies of system administrators touch on incident response but, 
largely due to the unpredictability of incidents, 
do so only briefly~\cite{barrett:cscw:2004,velasquez:2008,haber:2007}. In contrast we have designed our study to specifically capture (qualitative) data about multiple incidents, across multiple organizations.

One source of data we are using is incident reports, which are generally a product of \emph{postmortem investigations}, also called \emph{software failure investigations}. Various work considers how to effectively conduct such investigations, using controlled experiments~\cite{bjornson:2009} and other empirical methods~\cite{dalal2013empirical}, and propose associated process models~\cite{kohn2013integrated}. In our work we have not studied and have no visibility into how incident reports were created, but are simply using the result as done in other case study based research (eg.,~\cite{dalal2012case,eloff2018software}). 

Some aspects of our analysis has touched on support tools in use today, so it is worth highlighting some previous research related to (potentially) relevant tools. Note that this is necessarily an incomplete treatment. Detection and investigation activities often rely on state and behavior information logged by system components. Previous work has looked at tools and approaches for making more sophisticated use of such log data (e.g., automated log parsing~\cite{zhu:icse:2019}, log clustering tools for gaining insights into failures \cite{lin:icse:2016}, anomaly diagnosis through mining a time-weighted control flow graph in logs~\cite{jia:cloud:2017}). The visual representation of state and behavior information available to the responders in our study was quite simple (time based graphs of system metrics, say) and often not well integrated across information sources and types. Previous work in the area of both generic and domain specific software visualization may offer improvements (e.g.,~\cite{bodik:icac:2005,garduno:usenix:2012}).

\section{Study Setup}

The research we are reporting on proceeded in two phases. First, we interviewed engineers with relevant experience. From our interview data we have conducted an in-depth analysis of fifteen incidents and the associated incident responses. Second, we collected fifteen publicly published incident reports, with the goal of validating and refining the analytic results from phase one. These publicly published incident reports are typically the result of a \emph{postmortem analysis} of the incident conducted by the organization that owns the systems involved. The fifteen phase 1 incidents are referred to as \ii{1} to \ii{15} and the fifteen phase 2 incidents are referred to as \ip{1} to \ip{15}.

\subsection{Phase 1: Interview Study}

\subsubsection{Data Collection}

Our interview study involved eight participants, each with multiple years of experience responding to incidents (2, 5, 5, 2, 20, 20, 7 and 14 years respectively). Each interview was conducted by one or both of the authors of this paper and lasted approximately 60 minutes. During the interviews the participants were asked to describe particular incidents they were personally involved with. As the participants described these incidents the interviewers asked follow-up questions about both the incident and the response. Specifically, follow-up questions ensured we had details covering the cause of the incident, how it was detected, what the impact was, and how it was finally resolved. We also asked a series of questions (typically of the form ``what did you do next?'') covering the actions taken in response to each incident, and for each action we asked questions to capture \emph{who}, \emph{what}, \emph{how}, \emph{why} and \emph{with what result}. Each of the fifteen incidents are summarized in table~\ref{tbl:phase1}.


\subsubsection{Data Analysis}

We reviewed each of the fifteen incidents following a process similar to that described by Corbin and Strauss~\cite{corbin:2007}. As a first step, we used an open coding technique to identify individual properties of incidents. We identified these by comparing and contrasting the incidents and their associated properties, to identify which properties of incidents were distinguishing and relevant to differences we saw in incident responses. The process was iterative and interleaved with our data collection, and involved proposing and refining various categorizations as new incidents were encountered. Each of these categories are discussed in the results section of this paper (see section~\ref{results}).

We similarly analyzed each of the actions taken by the responders on a per incident basis. We iteratively developed a low-level categorization of incident response actions, identifying categories such as \emph{check recent code change} and \emph{scaling-up service}. We then looked across groups of related actions to identify investigative and mitigative strategies or patterns used by the responders. The low-level categories for actions are not reported in detail in this paper, but the strategies and patterns are discussed in sections~\ref{investigating} and~\ref{mitigating}.

\subsection{Phase 2: Publicly Documented Incidents}

\subsubsection{Data Collection}

To select fifteen incidents we followed a process that aimed to reduce bias in our collection (though see section \ref{validity} for a discussion of sources of bias we have not avoided) and also to find incidents with sufficient detail to support our analytic aims. First, we identified candidate lists of incidents using a Google search with the terms ``list of tech postmortems''. The search results included lists of incident reports and also lists of lists of such reports\footnote{E.g., \link{https://github.com/danluu/post-mortems}}. Second, we selected ten lists of incidents (directly and indirectly from the search results). All individual incident reports from the ten lists of incidents were added to a spreadsheet and their order was randomized. Next we applied predefined selection criteria, starting with the first incident (in the randomly sorted list) and stopping when 15 acceptable incidents had been identified. 

As selection criteria, we selected only incidents that occurred since 2015 (due to our interest in the current state of the practice), were in English, were publicly available and had sufficient information about aspects of the incident that was of analytic interest to us, namely information about how the incident (1) happened, (2) was detected, (3) investigated, and (4) mitigated. In this data collection process we briefly reviewed 584 incident reports, finding that most publicly published reports have details about the cause and mitigation of an incident (because the goal often is to reassure users that the issue has been resolved and will not reoccur) but are missing details about the detection and the investigation, making them unsuitable for our analysis. The 15 selected incidents are summarized in table~\ref{tbl:phase2}.

\subsubsection{Data Analysis}

We analyzed these fifteen incidents using the codes and categories defined in phase 1 of our research. So, rather than start with open coding, we began coding with an already defined coding scheme. However, we remained open to extending and refining our scheme based on this additional data. In the process we validated the generality of our scheme and extended it in several important ways. For example, we were able to further explore the concept of cascading failures in the context of the phase 2 incident reports.

Despite the selection criteria we used, when compared with phase one data, where we were able to ask follow-up questions, the publicly published incident reports we collected tend to have less detail about how investigations were carried out and do not always report on how the incident was initially detected. As a result, table~\ref{tbl:phase2} does not attempt to summarize the detection or the investigation of phase 2 incidents.

\subsection{Validity}\label{validity}

The concepts we have identified are grounded in the thirty incidents we have analyzed, and through our analysis we have refined them to the point that they are stable across those incidents. Towards the end of our analysis we were not adding new concepts, but refining and adding new examples of already discovered concepts. However, these concepts should not be viewed as comprehensive. Though our set of incidents is relatively diverse, it is possible (even probably) that with a different set of incidents other concepts would emerge.

Like most interview studies, the accuracy of the data from phase 1 of our work depends on the ability of our participants to recall details of events from the past. In addition, both data sets suffer from a related type of bias, which should be considered when interpreting our results. In phase 1, we found that our participants tended to share notable incidents rather than more routine ones, in some cases selecting memorable incidents that occurred some years earlier. Similarly, phase 2 data will be biased towards incidents that are significant enough to merit being published publicly. In both cases, we believe the consequence of this bias is that our data set is comprised of incidents of greater than average significance. However as we are not attempting statistical generalization nor attempting to capture the day to day work of engineers, the sample we have collected suits our research objectives.

\begin{table*}[t]
\centering
  \caption{Overview of phase 1 incidents.}
  \label{tbl:phase1}
  
  \begin{tabular}{
  r
  >{\raggedright\arraybackslash}p{2.35in}
  >{\raggedright\arraybackslash}p{0.45in}
  >{\raggedright\arraybackslash}p{2.35in}
  >{\raggedright\arraybackslash}p{0.95in}}
  
    \toprule
    {\bf \#} 
    & {\bf How it happened} 
    & {\bf Detection} 
    & {\bf Investigation} 
    & {\bf Mitigation \emph{(Time)}}
    \\ \midrule
    
    \ii{1}
    & Transient deployment failure led to nonfunctional web application \emph{(Deployment)}
    & Manual\newline Late
    & Assumed cause to be failed deployment \emph{(Opportunistic)}; no further investigation
    & Rollback to previous version \emph{(Hours)}
    \\ \addlinespace[.2em]

    \ii{2}
    & Deployments left previous application versions on server, eventually filling disk \emph{(Exceeding limits)}
    & Manual\newline Late
    & Investigated guesses (DNS, suspicious traffic, \ldots) and found the disk utilization \emph{(Opportunistic)}
    & Clean up files and restart \emph{(Hours)}
    \\ \addlinespace[.2em]

    \ii{3}
    & Database failure left a table locked, blocking subsequent transactions \emph{(System software)}
    & Auto\newline Late
    & Looked at transaction metrics and locks for correlated slow transactions \emph{(Opportunistic)}
    & Clean up locked transactions \emph{(Mins)}
    \\ \addlinespace[.2em]

    \ii{4}
    & Production usage exposed design flaw, leading to operations timing out \emph{(Exceeding limits)}
    & Manual\newline Late
    & Confirmed timeouts in logs; reproduced error at different layers to locate bottleneck \emph{(Systematic)}
    & Deploy fix making op. async \emph{(Days)}
    \\ \addlinespace[.2em]

    \ii{5}
    & Configuration change altered output data format of job, breaking downstream job \emph{(Deployment)}
    & Manual\newline Late
    & Traced failure from error logs to dependency \emph{(Systematic)} and to deployment \emph{(Opportunistic)}
    & Deploy fix and rerun job \emph{(Hours)}
    \\ \addlinespace[.2em]

    \ii{6}
    & Incoming rate exceeded capacity of data processing pipeline, leading to lost events \emph{(Exceeding limits)}
    & Manual\newline Late
    & Verified successive data processing stages to identify failing stage \emph{(Opportunistic+Systematic)}
    & Increase capacity of cluster \emph{(Week)}
    \\ \addlinespace[.2em]

    \ii{7}
    & Configuration change activated a new code path, exceeding capacity of datastore \emph{(Deployment)}
    & Auto\newline Late
    & Looked for correlated metrics, etc \emph{(Opportunistic)}; traced logged errors to cause \emph{(Systematic)}
    & Increase datastore capacity \emph{(Hours)}
    \\ \addlinespace[.2em]
         
    \ii{8}  
    & Deserialization defect led to gradually increasing number of cache misses and latency \emph{(Deployment)}
    & Auto\newline Late 
    & Correlated metrics and deployment \emph{(Opportunistic)}; investigated logs and code change \emph{(Systematic)}
    & Deploy fix for defect \emph{(Hours)}
    \\ \addlinespace[.2em]

    \ii{9}  
    & Growth exceeded allocated key size, leading to database transactions failing \emph{(Exceeding limits)}
    & Auto\newline On time 
    & Found no issues with metrics, etc \emph{(Opportunistic)}; reproduced error and checked logs \emph{(Systematic)}
    & Increase size of primary key \emph{(Mins)}
    \\ \addlinespace[.2em]

    \ii{10}
    & Router failure led to datacenter (and hosted systems) outage \emph{(System hardware)}
    & Auto\newline On time 
    & Noticed that systems were in the same datacenter \emph{(Opportunistic)}; verified unavailability \emph{(Systematic)}
    & Reinstall unavailable services \emph{(Hours)}
    \\ \addlinespace[.2em]

    \ii{11}
    & Configuration change in system A led to failures in A's calls to system B \emph{(Deployment)}
    & Auto\newline Late
    & Checked metrics, configuration changes, and logs \emph{(Opportunistic)}; traced errors \emph{(Systematic)}
    & Deploy configuration fix \emph{(Hours)}
    \\ \addlinespace[.2em]

    \ii{12}
    & Deployment increased logging rate leading to disks filling up and system failure \emph{(Deployment)}
    & Auto\newline Late
    & Checked metrics \emph{(Opportunistic)}; traced disk utilization to logs and code change \emph{(Systematic)}
    & Rollback and clean up logs \emph{(Hours)}
    \\ \addlinespace[.2em]

    \ii{13}
    & Accidental deployment of old configuration led to failure of multiple dependent services \emph{(Deployment)}
    & Auto\newline On time
    & Checked shared configurations and identified a recent configuration deployment \emph{(Opportunistic)}
    & Rollback and clear caches \emph{(Hours)}
    \\ \addlinespace[.2em]

    \ii{14}
    & Growth exceeded cache space, leading to evictions and intermittent missing content \emph{(Exceeding limits)}
    & Manual\newline Late
    & Checked for user error \emph{(Opportunistic)}; reproduced error by layer to isolate source \emph{(Systematic)}
    & Increase cache capacity \emph{(Days)}
    \\ \addlinespace[.2em]

    \ii{15}
    & Networking failure eliminated access to database, leading to multiple failures \emph{(System hardware)}
    & Auto\newline On time
    & Looked for and found in progress (company wide) issue that would explain all failures \emph{(Opportunistic)}
    & Restart systems once network up \emph{(Hours)}
    \\ \addlinespace[.1em]
    
    \bottomrule
  \end{tabular}
 \vspace{-0.2in}
\end{table*}

\section{Results}\label{results}

The results of our analysis are discussed here, organized into subsections covering: how the incidents occurred, how they were detected, how responders investigated and finally mitigated the incidents. The subsections first cover the categorization we developed in our analysis (which is also summarized by incident in tables~\ref{tbl:phase1} and \ref{tbl:phase2}), and then discuss the key observations we have identified. 

\subsection{How it happened}\label{hih}

A major theme identified in our analysis is that \emph{incidents grow in scope as an initial failure cascades through a system, exposing ways systems are not resilient to failure.} Sometimes this is as simple as when one node fails ``the remaining nodes would have to serve more consumers'' [\ip{9}] and are therefore more likely to fail; or when a database was experiencing issues due to higher than normal load, clients' ``attempted retries of failed writes [caused] elevated load'' and a database shutdown [\ip{5}]. In incident \ip{13} a brief network outage ``triggered a chain of events that led to 24 hours and 11 minutes of service degradation'' because it led the database cluster management software to structure the cluster in a way that was not supported by the client applications and the data replication processes. Despite these cascading effects, it is generally possible to identify a precipitating event for the incident, and we have categorized these events as follows.

%
%
\category{Deployments} Seven of our phase 1 incidents and three of our phase 2 incidents were precipitated by a code or configuration deployment. Two of the seven phase 1 incidents precipitated by deployments were exceptional. First, the deployment that led to incident \ii{13} was an \emph{accidental} deployment of a set of 12 month old configurations. Second, incident \ii{1}, was not caused by a deployment of a defect, but rather the deployment itself failed, leaving the application in a half-installed, non-functional state.

%
%
\category{Infrastructure change} Three of our phase 2 incidents were precipitated by an infrastructure change or routine maintenance that had unexpected consequences. For example, incident \ip{3} began when the system's primary datastore instance was automatically replaced and the system was unable to properly failover to the backup instance due to a preexisting, known defect. 

%
%
\category{Exceeding scaling limits} Five of our phase 1 incidents and six of our phase 2 incidents were precipitated by usage growth beyond some scaling limits and experiencing issues ``only revealed at [the new] peak production workloads'' [\ip{8}]. As a simple example, in incident \ii{9} a database key size limit was exceeded causing database transaction failures. 

Similarly, a system may exhaust available resources, which implies a resource leak, the absence of appropriate resource management functionality, or simply a failure to allocate sufficient resources to begin with. For example, incident \ii{2} occurred because each application install left the previous version on the server, eventually filling up the hard drive of the host machines. A previous team member had been manually running a script to clean up the application instances, but when he left the team, that task was discontinued due to ignorance. Incidents \ii{7} and \ii{12} were categorized as being caused by a \emph{deployment}, however those deployments led to limits being exceeded and so they could have been included here.


%
%
\category{System software or hardware failure} Three of our phase 1 incidents and three of our phase 2 incidents were precipitated by a failure in underlying system software or hardware. Such failure events trigger incidents when the applications lack the ability to tolerate such failures. For example, in incident \ip{15}, a single cluster instance experienced a CPU locking event and the cluster management software failed to successfully remove that instance from the cluster.


Nearly all of the incidents we have analyzed represent, at least partially, a failure in testing (also identified in previous work, such as~\cite{dalal2012case}), combined with an inability to tolerate various failure states. The following four key observations shed some light on the challenges and limitations of testing as it is practiced today.

%
%
\obs{obs:cases}{Test cases often fail to detect defects that lead to incidents only when (possibly rare) combinations of events or system states coincide.}
When combinations of things going wrong are needed to trigger a failure/incident often that combination may not be found in test suites, given the complexity and subtlety of the interactions involved. These challenges (at least partially) explain why preexisting defects can remain latent and why some defects make it to production in the first place (see also~\cite{controneo:issre:2013}).

For the same reason, new defects can be introduced even in code bases with thorough tests suites. In fact, our analysis of incident \ip{10} found that ``three conditions \ldots were necessary'' for the incident: ``a misconfiguration that only surfaces when two processes crash at almost the same time.'' A combination unlikely to arise in typical testing strategies because it is unlikely to be anticipated.

Incident \ip{2} was the result of a combination of defects: one a preexisting, unknown code defect in a shared framework that left a per-request variable referenced after the request was complete; and the other a newly introduced configuration defect in a client of that framework, that set the request timeout limit to a larger than intended value. The combination of these two defects was needed to cause the memory leak that led to the incident.

%
%
\obs{obs:environment}{Testing environments and other preproduction environments often do not capture all aspects of the production environment.}
In addition to the inability for test cases to capture all possible combinations of events (as just discussed), another limitation of testing we observed in our analysis is that the environments that new code and configuration run in before reaching production do not capture all aspects of the production environment. So the issue may not be that the appropriate test cases are missing, instead it is that the tests are being run in a context where it passes, even if there is an issue that would manifest in production because test environments are necessarily simplified in various ways. As a simple example, a cache deserialization defect caused incident \ii{8} and was not caught in testing because the test environment was configure only with a local cache which did not exercise the deserialization code, while production used a distributed cache.

Both observation~\ref{obs:cases} and~\ref{obs:environment} relate to previous work about software failure reproducibility. For example, Caveezza, et al.  identified environmental factors that influence reproducibility: memory occupancy, disk usage and concurrency level~\cite{caveezza:sre:2014}. Complementing this previous work, we have identified three general ways that test environments differ from production environments. (1) Test environments may not be in a ``position to replicate the volume of traffic'' [\ii{14}] or otherwise run at production scale. For example, incident \ip{2} occurred when the system's database setup was changed, which change looked fine until the daily peak period was reached and it could not keep up with the transaction rate, causing ``a cascade of other issues.'' (2) New code may not be run in test environments long enough to expose some resource exhaustion issues. For example, in incident \ii{12}, the deployment of new logging code increased the rate of logging, resulting in insufficient disk space after a period of time that was longer than the time spent in testing. (3) Test environments do not always exercise code in all non-typical (i.e., error) states experienced in production. For example, in incident \ii{15} once network connectivity was restored, some of the affected services did not return to normal operation and needed to be restarted.

%
%
\obs{obs:limits}{When scaling limits are not well known, tested or monitored for they are discovered when they are exceeded.}
In addition to failing to capture the scale of the current production environment as just discussed, failure to imagine how future growth would impact system behavior and test accordingly precipitated multiple incidents. Organizations do not always run tests that let them know at what scale their system will fail and so it fails unexpectedly. Even post incident, after scaling up the system in some way, it is not always clear ``how long are we rebuying ourselves?'' [\ii{14}] 

Many tests are static and do not change as the production environment changes. See, for example, incidents \ii{6} and \ii{14} where systems that passed tests eventually failed in production as usage grew. In neither case were the teams aware of, or paying attention to, the relevant limits (ingestion cluster capacity and cache space, respectively). In the case of incident \ip{9} the (OS or hardware) limit on outbound network traffic for machines in a data processing cluster was exceeded and the team had not been monitoring that aspect of scaling, instead ``paying closer attention to CPU and disk.'' It was easy to tell what CPU and disk limits needed to be avoided, but the maximum outbound network limit was unknown and ``typically depends on a mix of factors.'' 

%
%
\obs{obs:config}{Configuration changes are just as risky as code changes, but are often not tested and deployed with the same care.}
For high availability systems, source code changes tend to move through a reviewing, testing and gradual deployment process not always mirrored for configuration changes. For example, in incident \ii{11} there was a defective configuration change with no associated deployment. 

Incident \ip{12} was precipitated by a configuration change that led to a global outage of multiple services. In this organization, configuration changes go through a short test and deployment cycle that can ``push changes globally in seconds'', where code goes through a longer process that ``can take hours or days'', partially because it proceeds one region at a time, reducing the potential scope of some types of incidents that might arise. Note that in the context of responding to security incidents, a quick and global configuration deployment is an advantage, and their ``customers have come to love this high speed configurability.''


\begin{table*}[t]
\centering
  \caption{Overview of phase 2 incidents.}
  \label{tbl:phase2}
  \begin{tabular}{
    r
    l
    >{\raggedright\arraybackslash}p{3.1in}
    >{\raggedright\arraybackslash}p{1.4in}
    l
    l}
    
    \toprule
    {\bf \#} & 
    {\bf Report} &
    {\bf How it happened} & 
    {\bf Mitigation} & 
    {\bf Caught} &
    {\bf Time}
    \\ \midrule
    
    \ip{1} 
    & \link{bit.ly/2xTn1iK}
    & Database downgrade and connection pool configuration change led to outage when daily peak load reached \emph{(Infrastructure change)}
    & Multistep rollback of setup, increasing capacity
    & Late & Hours
    \\ \addlinespace[.2em]
    
    \ip{2} 
    & \link{bit.ly/2UNWBrI}
    & Deployment of a configuration defect + a preexisting code defect led to a memory leak and a service outage \emph{(Deployment)}
    & Deploy fix for new and preexisting defects
    & Late & Hours
    \\ \addlinespace[.2em]

    \ip{3} 
    & \link{bit.ly/2Vku38n}
    & Automated migration of primary datastore instance + a preexisting failover defect led to cascading outages \emph{(Infrastructure change)}
    & A full restart of the service (after partial restarts failed)
    & On time & Hours
    \\ \addlinespace[.2em]
    
    \ip{4}
    & \link{bit.ly/3aY3Hj5}
    & Deployment altered webserver's buffering strategy, exposed a preexisting defect, and led to a buffer overflow \emph{(Deployment)}
    & Deploy fix, clear caches, and reenable features
    & Late & Hours
    \\ \addlinespace[.2em]

    \ip{5}
    & \link{bit.ly/2yPjFOl}
    & Database maintenance processes failed due high load so transaction IDs exceeded limit and database shutdown \emph{(Exceeding limits)}
    & Clean up data allowing process to complete
    & Late & Days
    \\ \addlinespace[.2em]
        
    \ip{6}
    & \link{bit.ly/2wpqMMq}
    & Database replication failure due to high load and and accidental deletion of data led to database outage \emph{(Exceeding limits)}
    & Partially restore deleted data and reenable features
    & On time & Hours
    \\ \addlinespace[.2em]

    \ip{7}
    & \link{bit.ly/3b3N6uo}
    & Exceeding disk space and available throughput limits led to a failed deployment and cluster instances in bad states \emph{(Exceeding limits)}
    & Restart instances by terminating them 
    & Late & Hours
    \\ \addlinespace[.2em]
    
    \ip{8}
    & \link{bit.ly/2XqzFjR}
    & High traffic caused 6 outages due to exceeding various database, loadbalancer, memcached limits \emph{(Exceeding limits)}
    & Manually failover database, increase capacities and limits
    & On time & Hours
    \\ \addlinespace[.2em]
    
    \ip{9} 
    & \link{bit.ly/3c7sFwq}
    & Usage growth exceeded network capacity of nodes in a data processing cluster and led to a cascading outage \emph{(Exceeding limits)}
    & Increase cluster capacity and deploy fixed configuration
    & On time & Hours
    \\ \addlinespace[.2em]

    \ip{10} 
    & \link{bit.ly/39UxK9R}
    & Cluster management misconfiguration + coincident disk failure and process crash led to outage \emph{(System software or hardware failure)}
    & Manually failover and restart cluster
    & On time & Hours
    \\ \addlinespace[.2em]

    \ip{11} 
    & \link{bit.ly/39Z5dQu}
    & Existing configuration defect + multiple database failures caused a failed failover and an outage \emph{(System software or hardware failure)}
    & Rollout fixed configuration and restart all nodes in cluster
    & On time & Hours
    \\ \addlinespace[.2em]
    
    \ip{12}
    & \link{bit.ly/2RoVfBu}
    & Configuration deployment led to an increase in CPU usage and a global outage of multiple services \emph{(Deployment)}
    & Rollback configuration and reenable features
    & On time & Mins
    \\ \addlinespace[.2em]

    \ip{13} 
    & \link{bit.ly/3aYpKGq}
    & Routine network interruption led to misconfigured database clusters, affecting latency and availability \emph{(Infrastructure change)}
    & Restore missing data and manually failover database
    & On time & Day
    \\ \addlinespace[.2em]

    \ip{14} 
    & \link{bit.ly/2yRMolL}
    & Errors in a dependency caused excessive logging and full disks, which crashed application on same servers \emph{(Exceeding limits)}
    & Clean up logged data and restart application
    & Late & Hours
    \\ \addlinespace[.2em]
    
    \ip{15} 
    & \link{bit.ly/34qRt03}
    & Errors on one cluster instance + a preexisting defect led to a split cluster and delays for dependent clusters \emph{(System software)}
    & Multistep restart of entire service and reenable features
    & On time & Mins
    \\ \addlinespace[.1em]
    
    \bottomrule
  \end{tabular}
   \vspace{-0.2in}
\end{table*}

\subsection{Detecting}\label{monitoring}

Incident response begins when some abnormal behavior (i.e., one or more initial symptoms) is observed, either automatically or manually, as summarized in table~\ref{tbl:phase1} for phase 1 incidents. For phase 2 whether the detection was automated or manual was not always specified in the incident reports we analyzed. Automatic detection is based on a set of support systems that we will collectively refer to as monitoring and notification systems. The monitoring systems our participants used tended to include: hardware and system level metrics (e.g., disk utilization), application metrics (e.g., latency per request), application logs (e.g., error logs), and a few other miscellaneous items such as job statuses and deployment histories. 

Monitoring (in production environments) compliments testing, overcoming some of the limitations of testing we just discussed. A defect that is missed by testing (say, because the correct combination of events are not covered in the test suite, as discussed in observation \ref{obs:cases} above), can be caught by a monitoring system, if and when it is triggered. Unlike tests, which must attempt to recreate a production like environment (see observation~\ref{obs:environment}), monitoring can verify that various properties and behaviors hold in production. Finally, even if scaling limits have not been identified (see observation~\ref{obs:limits}), a monitoring system can detect failures related to scaling, using proxy metrics.

In this section we discuss our analytic results related to monitoring and detecting from three perspectives or dimensions: \emph{automation}, \emph{specificity}, and \emph{timeliness}. The ideal detection is automated, timely and provides a specific starting point for the investigation. Incidents that are discovered late, only discovered accidentally or begin from a generic starting point are analytically interesting as they potentially indicate challenges or limitations with monitoring as commonly practiced today. 

%
%
\category{Automation} We categorized our phase 1 incidents into those that were automatically detected and those that were manually detected. When detection was manual, it was generally accidentally discovered by someone associated with the team that owned the system or was reported by user of the system. Manual detection can also occur as engineers inspect logged system messages, though such signals can be missed simply due to the volume of information (for example, ``unrelated exceptions coming in from other systems drowned it out [and delayed] detection'' of incident~\ip{5}) or ``shrugged off as a transient issue that would resolve itself'' [\ip{7}].

%
%
\category{Specificity} The starting point (i.e., the initial symptom or abnormal behavior) for an investigation could be more or less specific to the root cause of the incident. A less specific (or more generic) symptom is one that could have many possible root causes and the causal chain from root cause to the symptom may have multiple steps. Incident \ip{2} began with ``spiky and mysterious system metrics'' followed by a multi-hour ``process of uncovering the root cause.''

%
%
\category{Timeliness} Building on the previous point, many generic metrics are trailing metrics and a more specific metric might be more timely. In the third column of table~\ref{tbl:phase1} and the second last column of table~\ref{tbl:phase2}, we categorize each incident based on whether it was detected \emph{on time} (meaning it was detected roughly as soon as the issue began) or \emph{late}. We would consider a detection \emph{early} if it prevents the incident from occurring, but we have no such examples in our data since we have only analyzed incidents that did occur. Unsurprisingly, all of the manually detected incidents were detected late.

%
%
\obs{obs:catchall}{Generic catch all monitoring and notification are important but tend to be trailing metrics, leading to late detection.}
We have not attempted to quantify the specificity of each of the initial symptoms in our data set, however none of the initial notifications in either phase of our study were highly specific, even when detection was automated. This is partially because some detection is based on \emph{catch-all} notifications, which are important because they are able to catch incidents arising from unanticipated root causes. However, with such catch-all notifications, the starting point is generic (roughly, ``something is wrong with your system or one of its dependencies'') and further investigation tends to be needed before a mitigation strategy can be pursued. 


For some incidents it is possible to argue that some monitoring is missing and would have led to earlier detection. For example, in incident \ii{8}, error messages in the logs (about failures to deserialize cached objects) began to appear several hours before the latency metric exceeded the configured threshold and triggered the notification. Similarly, in incident \ii{12} a disk utilization based notification (rather than the latency metric, described by the responder as ``pretty generic'') would have led to more timely and specific notification.

%
%
\obs{obs:thresholds}{Pre-determined, threshold based detection is fragile and incomplete.}
Automated detection is typically based on two things: system monitoring and rule-based notifications. Automated monitoring can be: (1) white-box monitoring, based on event and metric logging generated by the system being monitored, provides an internal view into the state and functioning of the system; or (2) black-box monitoring based on the results of an external system interacting with the system being monitored. Rule based notifications are built on top of such monitoring systems. These rules can be event or threshold based. Since many system's monitoring data is noisy, tuning these requires care to avoid false positives or negatives, and is an on going and largely manual activity in practice. 

Naturally, foresight is limited. For incident \ip{2}, the team was not notified of memory usage growth because appropriate notifications were not put in place. As incidents are reviewed retrospectively, monitoring is added, removed and ``retuned'' to improve it along the three dimensions we have discussed here. For example, during the mitigation of incident \ii{6} a monitor was put in place to detect dropped events to help verify that the mitigation was successful and help prevent future incidents. Over time, teams build up large sets of metrics and associated threshold based notifications: ``we got metrics \ldots for things like latency, errors, dropped connections, you got your memory usage, hard drive space, you got every hardware metric that you may think of'' [\ii{12}]. The ongoing maintenance effort (as requirements change, for example) involved is significant and yet still the notifications remain incomplete. 

%
%
\obs{obs:tested}{Monitoring, notification and other support systems may not themselves be as well tested or monitored as primary systems.}
Based on our analysis it appears that system support tools, are not as well tested or monitored as primary systems and often lack redundancy, and yet defects in these systems can affect all aspects of incident detection and response. A simple example is that notification for incident \ip{1} was delayed four hours ``due to misconfigured settings'' in their notification tool, and the responders also realized that, the process they intended to use as part of the mitigation, works ``really well if [their service] is online'' but not otherwise.

The database backup procedure used by the organization involved in incident \ip{6} was failing for some time before the incident, due to a version mismatch between the database and the backup tool, however the team was ``never aware of the backups failing'' because the email notification system was misconfigured, until they needed the backups to mitigate the incident and ``there was no recent backup to be found anywhere.''

\subsection{Investigating}\label{investigating}

The goal of incident response is to mitigate the issue, given some initial abnormal behavior as a starting point. When the starting point for the incident response was a generic symptom (one that was distant from the root cause, say) as discussed in section~\ref{monitoring}, mitigating an incident first required an investigation into the cause. Responders used different information gathering activities, following either a \emph{systematic} or \emph{opportunistic} strategy, or a combination of the two. Incident \ii{1} is exceptional in that a correlation with a recent deployment was assumed, but no further (opportunistic or systematic) investigation was done. The deployment was successfully rolled-back and verified, suggesting the failed deployment was in fact the root cause for the incident, as assumed.

%
%
\emph{Systematic strategies} start from a symptom and follow a step-by-step investigation to the root cause, following two related approaches. (1) \emph{Tracing a chain of behaviors or symptoms}. This approach involves asking a series of \emph{why} questions until a root cause is discovered. (2) \emph{Stepping through the technology stack}, reproducing the issue at different layers or stepping through a series of dependent systems until a point of origin for the issue is identified: ``you just walk your way down the stack; when you find places where we can reproduce the issue you just keep walking down the stack until you find the place where it really exists'' [\ii{14}].

%
%
\emph{Opportunistic strategies} are ways to conceptually jump directly to a (possible) root cause or at least finding other symptoms closer to the root cause, without tracing step-by-step from the initial abnormal behavior to that cause. Our participants followed one of two approaches in applying this strategy. (1) \emph{Checking common (or typical) causes for the abnormal behavior}. Such investigations rely on past experience and often represent \emph{low-hanging fruit} in that they can be checked without an extensive investigation. (2) \emph{Looking for anomalies that are temporally correlated} with the behavior under investigation. A detailed causal chain between the correlated items may be filled-in later, or just assumed without additional investigation. 

We have found that an initial opportunistic strategy may be used to provide the responder(s) with a more specific point to start a systematic investigation, and is strongly organized around the concept of an \emph{incident timeline}: ``the timing [of the symptoms] is very important because it allows you to correlate'' across information sources [\ii{3}]. When an incident begins with multiple notifications, the responder may begin by attempting to determine what is ``the point of commonality'' [\ii{10}], or in other words, asking what do the notifications have in common, since coincident notifications likely share a common root cause. In the case of incident \ii{10}, the responder determined that all of the affected systems were in the same data center, allowing him to quickly identify the root cause.

Table~\ref{tbl:phase1} shows which strategies were used in each of our fifteen phase 1 incidents. As noted earlier, the phase 2 incident reports did not always have investigation details, however the the details that were available (from both phases) helped inform the following two key observations.

%
%
\obs{obs:opportunistic}{Opportunistic investigative strategies are successful if correlations are surfaced and understood.}
In looking for correlations, the same sources of information used by monitoring systems, as described in the previous section, were used by the responders in their investigations. In the best cases, the information needed to draw these (often temporal) correlations is accessible in tools like system dashboards, which present time based graphs of metrics and events. Visually scanning such graphs was enough for the responder from \ii{7} to quickly determine the root cause and collect additional relevant information. When the needed information was not so readily available and integrated into dashboards, investigation was cognitively more difficult. Responders also used less accessible, manually extracted information as needed. For example, responders may search through logs (\ii{7}, \ip{14}), read through documentation (\ii{10}, \ip{2}), sample instance state (\ii{15}, \ip{7}), or collect and review memory dumps (\ii{12}, \ip{2}).

During incident \ii{11}, the responders identified a correlated configuration change but failed to understand the causal relationship to the incident. The team members did not understand the configuration and associated systems and so they did not investigate it further. Hours later, a second team of responders successfully recognized that a defect in the configuration change was the root cause. Similarly, the responders to incident \ip{2} saw that a configuration change ``had a very good correlation in terms of timing'' but when reviewing that change they failed  to see the relationship with the memory leak. Their investigation then proceeded systematically for several hours using memory analysis tools, ``some of which were more fruitful than others'' [\ip{2}], before the relationship between the configuration change and the event was understood.

%
%
\obs{obs:architecture}{Architectural complexity has an operational (including investigative) cost.}
In our analysis we identified four types or sources of architectural complexity that make incident response, and investigation specifically, more difficult. (1) Architectural ``layering'' made it difficult to determine at what layer the issue existed and where to focus mitigation efforts. For example, incident \ip{7} involved a cluster of machines, each with applications running in containers, managed by a container/cluster management tool, and the responder struggled to identify at which layer the event was occurring. (2) Nondeterminism can defeat investigation strategies that rely on reproducing errors. For example, during incident \ii{14} nondeterministic cache evictions meant that ``hosts didn't behave the same way.'' Similarly, a previous study of space mission faults found that 36.5\% were not deterministically reproducible~\cite{grottke:ifip:2010}. (3) A component with multiple dependencies has multiple points of failure (``most of the time when a latency ticket occurs it is based on one of our dependencies'' [\ii{12}]). Of course, the ideal would be to engineer components to be tolerant to failures in dependencies, however this is not always the case in practice. (4) Failover algorithms for clusters and replicated database systems made it difficult to understand and mitigate incidents because their decision making processes are difficult to predict and they can arrive at surprising states. A key issue in the cascading failures causing incident \ip{13} was a cluster management system configuring a cluster's topology in a way that was not predicted or tested for.

During initial incident investigations, the responders were not always able to completely determine the root cause of the incident, further illustrating the system/incident complexity we are highlighting here. Later during a postmortem analysis, where there is more time and all available information can be reviewed, a more complete understanding could be developed. Though responders to incident \ip{5} identified two possible explanations for how an incident happened but were ``unable to determine for certain which occurred'' even after a postmortem analysis.

\subsection{Mitigating}\label{mitigating}

In this study, we consider an incident response complete once it has been mitigated and that mitigation has been verified by the responders. All mitigations that we analyzed appeared to address the immediate root cause of the issue, with the exception of incident \ii{1}, as discussed above. In some cases this mitigation involved multiple steps and took hours or even days to complete. During incident \ip{1}, responders ``made some temporary fixes to get [the service] back online'', though response times and availability remained a problem until mitigation was complete, about one hour later. At times the mitigation was gradual (``by half hour you'll see that [it is] pretty good, and a hundred percent by one hour'' [\ii{13}]).

Though we are discussing investigating and mitigating separately in this paper, they tended to be intertwined in interesting and important ways. Investigation focuses on how to mitigate and the results of mitigation activities (including failed mitigations) inform the ongoing investigation. We also observed that the verification closely mirrors the investigation. Once a mitigation action was taken, responders verified that the issue was resolved by checking that the relevant symptoms (from the original notification and subsequent investigation) returned to normal over a timeline that would be expected given the responder's understanding of the root cause; including checking that the original ``indicators that fired the [notification\ldots] started turning to normal'' [\ip{15}]. 


Our analysis identified five (non-mutually exclusive) mitigation strategies used by responders. Each of these strategies are discussed below and our key observations follow that discussion. See tables~\ref{tbl:phase1} and~\ref{tbl:phase2} for a summary of which mitigation strategies were used as part of which incident responses.

%
%
\category{Rollback} When the incident is precipitated by a deployment, rolling back to a previous version of the system is a natural mitigation in some cases, and was part of resolving three phase 1 incidents and two phase 2 incidents. Early on in incident \ii{8}, the responder attempted a rollback, since he (rightly) believed that a defect related to writing to a cache had been deployed. However, the rollback attempt was abandoned because the (bad) state of the cache persisted and made it appear that the previous version was also defective, and the responder was not confident enough to ignore the errors and complete the rollback.

%
%
\category{Deploy fix} When a defect has been deployed, an alternative to rolling back to a previous version is to fix the (configuration or code) defect and deploy that fix, as done in four phase 1 incidents and three phase 2 incidents. In some cases, this strategy can be seen as a partial rollback targeting just the deployed defect, leaving other deployed changes in place.

Amongst incidents precipitated by a deployment, we found that deploying a fix was slightly more common than rolling back the deployment. We identified two reasons for this preference. (1) A rollback is a blunt tool, possibly rolling back more than just the defect that is the root cause.  In incident \ii{12}, a change that increased the rate of logging was the only change made in a particular deployment so it would be ``safe and easy to rollback'' however if ``there were a bunch of other commits, then it would be more difficult.'' (2) A rollback does not restore all aspects of the (persistent) state of the system and therefore may be insufficient mitigation (see observation \ref{obs:mitigating}). During incident \ip{2} the responders, opted not to rollback the precipitating deployment because they could not see how it would have caused the incident. With the benefit of hindsight, the responders wished they had rolled-back the change which ``would have fixed it and given information about root cause.''

%
%
\category{Increase capacity} Mitigating four phase 1 incidents and three phase 2 incidents included allocating additional resources, because some operating limit had been exceeded. For example, incident \ii{7} began with a configuration change which ``turned on'' a new code path, sending more traffic to an under-provisioned backend service, and was mitigated by allocating additional backend capacity.

%
%
\category{Clean up or restore resources} Two phase 1 incidents and four phase 2 incidents involved cleaning up resources, such as restoring data from backups. For incident \ii{12}, in addition to deploying a fix for the excessive logging, the responder also deleted large log files to complete the mitigation. In response to incident \ip{13} the responders manually reconciled over 900 database writes that had failed to replicate. Additionally, mitigating three phase 2 incidents involved manually performing a failover operation, which is a specialized restoration that changes which instance is primary in a (database or other) cluster. The operation was more or less straightforward depending on how well data replication had performed over the course of the incident. During incident \ip{8}, the responders were performing a ``manual primary failover [\ldots] multiple times per hour'' until the incident was completely resolved.

%
%
\category{Restarting, reinstalling, reenabling} 
Related to cleaning up resources, when clusters, software, machines, or other resources end up in an error state as part of an incident, restarting or reinstalling those can result in resources being restored to a healthy state, which played a role in mitigating two phase 1 incidents and seven phase 2 incidents. Restarting can take the form of deleting instances (for example cluster instances as in incident \ip{7}) which are automatically replaced. However, a restart may temporarily make the impact worse. For example, the responders in incident \ip{3} were ``forced to fully restart the service'' due to the large number of components in a bad (not fully understood) state, temporarily disconnecting millions of concurrent users. Similarly, responders in incident \ip{15} first deployed patches to ``introduce limitations on connections [\ldots] to assist in rebooting.''

We observed that in some cases responders turned off or disabled features (``in an attempt to shed load'' [\ip{15}], say) and  mitigating the incident involved reenabling those features. Some systems had such on-off switches for features, but during the response to incident \ip{4}, the responders ``implemented a global kill'' for an impacted feature. Since database writes were not replicating correctly, while responding to incident \ip{13} the responders deliberately turned off service features. Their ``strategy was to prioritize data integrity over site usability and time to recover.''


%
%
\obs{obs:mitigating}{Addressing the original root cause is not always sufficient mitigation; the mitigation does not always cascade the way the failure did.}
We observed that there are two distinct kinds of problems that need to be considered and resolved during mitigation: (1) the original problem that led to the incident, and (2) cascading problems that came later, including systems getting into error states and problems cascading between components. For example, responders in incident \ip{4} needed to ``remove any cached HTTP responses'' in addition to fixing the defect that was causing corrupted HTTP responses. During incident \ip{3} responders found that after resolving the initial ``partial outage, unnoticed issues on other services caused a cascading failure'' that also needed mitigation. When a ``failure [leads] to a number of cascading failures'' it can take ``time to fully unwind'' [\ip{9}]. Incidents that result in data inconsistency or data processing issues (un-replicated database writes, say) can have particularly time consuming mitigations. 

 

%
%
\obs{obs:careful}{Changes made in the context of incident response have the potential to make issues worse (and are made with fewer precautions than typical).}
As discussed above, a restart of a service or other resource may temporarily make an incident worse, but were undertaken deliberately as part of mitigating the incident. We also saw instances where attempted mitigation actions \emph{unexpectedly} made matters worse. Most mitigation actions in our data set were taken with little testing due to the urgency of the issue, except during incident \ii{14} where responders were dealing with a partial outage and tested the planned mitigation in a non-production context and waited until non-peak hours to make the changes to ensure they did not ``make things worse.''

Hurried interventions are undertaken that are not common (well practiced) operational procedures. In many cases it seems that this is appropriate. An activity that might lead to an outage, could appropriately be attempted if there is already an outage. However, as responders to \ip{11} found, a further failure can occur as a ``result of our remediation efforts.'' In this case a rollback of a configuration defect lead to a combination of configurations that had never been tested together and were incompatible. During the response to incident \ip{6}, one responder, attempting to speed up a data replication process, accidentally deleted ``around 300 GBs of data'' some of which was unrecoverable.

The same risks exists for automated mitigations. For example, during incident \ip{8} the service was experiencing slow response times and a partial outage. Due to this partial outage, the machines running the service (which themselves were healthy) began failing automated ``health check requests [causing] the load balancer to pull all the nodes out of rotation imposing full service downtime.''


%



\section{Implications}\label{implications}

Though the focus of our data collection and analysis has been on the state of the practice, our categorized analytic results and \ref{obs:careful} key observations have relevance to multiple areas of research. For example, while fault tolerance is not a new idea (e.g., \cite{pierce1965failure, hecht1976fault}), our analysis suggests that the state of the practice is far from ideal and a major theme identified in our analysis is that failures can cascade, making the difference between brief, local incidents and major outages. A common lesson learned from postmortem reviews is ``to think harder about cascading failure scenarios'' [\ip{9}]. Similarly, testing and verification are mature research fields (e.g.,~\cite{clarke:toplas:1986,rapps:tse:1985}), but challenges around creating tests and testing environments persist (see observations~\ref{obs:cases} and~\ref{obs:environment}). To concretely illustrate a few of the implications of our results, we conclude this paper with a brief discussion of three suggested directions for follow on work.


\impl{Inferring System State}

As the scale and complexity of the systems under maintenance increases, manual management of notifications becomes less feasible (see observations~\ref{obs:catchall} and~\ref{obs:thresholds}). Though not generally used in the incidents we have analyzed, multiple anomaly detection methods based on automated log analysis have been proposed (see for example~\cite{lin:icse:2016,du:ccs:2017}; and see~\cite{he:issre:2016} for an evaluation of multiple techniques) and may eventually replace manually configured notifications. First, improvements to the state of the art will need to include improved running time to support real-time analysis of large logs, improved accuracy (especially improved recall), and a focus on anomalous metric values, rather than just boolean events. 


Much of the previous anomaly detection research takes a generic (i.e., not application specific) approach and simply identifies an event as anomalous. However, the monitoring ideal may well be a system that can accurately infer the state of systems and components, even if it needed to be application specific. To this end, such inference systems would need to understand or infer the relationships between components and need to correlate metrics and events across components. A state inference system could be foundational for multiple support use cases. As a simple example, the rule that lead to the notification for incident \ii{12} is roughly configured as ``if three 99.9 percentile consecutive data points over 200ms are observed then create a new notification.'' With an inference system, the improved version could be ``if latency is consistently or intermittently elevated create a new notification'', requiring less ongoing maintenance. 

\impl{Automated Mitigation}


Automated mitigation systems have the possibility of mitigating incidents more quickly than a human can respond, though mitigating actions also have the potential to make the incident worse (see observation~\ref{obs:careful}). Automated mitigation is related broadly to research in the area of \emph{self-healing} systems~\cite{schneider:spae:2015} and Autonomic Computing~\cite{horn:2001}, both of which broadly aim to create system support software that can autonomously manage systems. Improved state inference, as just discussed, could help move us toward more fully autonomous mitigation by allowing engineers to program rule based mitigative actions. While exploring this possibility in detail is outside the scope of this paper, for the sake of concreteness here is a set of possible inferred states along with possible automated actions that could be taken based on those states. Each of these examples are related to incidents in our data and we imagine that a broader data set of incidents could lead to a much larger set of examples. 

\begin{enumerate}
  \item If \emph{performance is degraded due to dependency D}, then \emph{increase the TTL for cache of data from D}.
  \item If \emph{disk utilization is high and growing across fleet}, then \emph{increase frequency of log rotation and deletion.} 
  \item If \emph{fleet's CPU load is high due to temporary, high load}, then \emph{reduce maximum concurrent connections.}
  \item If \emph{fleet is under memory pressure due to memory leak}, then \emph{schedule rolling reboot of fleet, 20\% at a time}
\end{enumerate}

One type of automated mitigation relevant to our data (see observation~\ref{obs:limits}) is autoscaling, which is built into many cloud computing resources (see~\cite{botran:grid:2014} for a summary of autoscaling approaches). We argue that autoscaling could be improved if it could be based on more (and more accurate) information about the state of the system: what is the scaling limit that is being exceeded (or nearly exceeded)? What effect is the ongoing autoscaling having on the state of the system? There are cases where even if a scaling limit is reached, adding resources (e.g., adding hosts to a fleet) will not adress the issue (because a dependency may be overwhelmed, say) or may not be desirable (the increase in traffic is temporary and scaling up would be cost prohibitive) in which case a load-shedding strategy maybe more appropriate. 

\impl{Incident Specific Dashboards}

System dashboards played a helpful role in investigating incidents, allowing responders (in some cases) to identify other symptoms that were more specific or closer to the root cause than the symptom that lead to the initial notification. Dashboards as used by our responders displayed charts with pre-selected set of metrics, and were necessarily incomplete, covering a particular system and particular information types and sources. We observed cases where information existed (in say a log file on a particular host) but was not readily accessible and therefore only discovered after a more time consuming and systematic investigation. 


More useful to responders would be an automatically generated dashboard that is \emph{incident specific}, and integrates and correlates across a broad range of systems and information sources. An algorithm that integrates and correlates anomalous data with an initial symptom would be the minimal level of support needed (see~\cite{mattern:1988}). Both the correlation and information display should be based strongly on the concept of an incident timeline. Beyond correlations and visualizations of metrics over time, more sophisticated inference and execution visualizations may reduce the cognitive effort required of engineers during investigation (e.g.,~\cite{beschastnikh:icse:2014,beschastnikh:acm:2016}).

\section{Summary}

In this paper we have reported on the state of the practice for how incidents occur, are detected, investigated and mitigated, based on an analysis of a set of thirty incidents. We have highlighted the challenges engineers face in constructing tests suites and test environments suitable for detecting code and configuration defects that lead to incidents; their difficulties predicting, testing and monitoring the way that systems behave as scale increases; the cost of manually maintaining monitoring systems, deficiencies in which can lead to late or missed detections; the cognitive challenges of investigating incidents in the context of architectural complexity and diagnostic information scattered across different support systems; and finally the way that cascading faults increase the impact of incidents and complicate mitigation. We hope these insights will lead to improvements to how we evolve, maintain and support software systems in production.

\clearpage 
\bibliography{references}{}
\bibliographystyle{plain}

\end{document}